# Medium-sized $Si_n^-$ (*n*=14–20) clusters: a combined study of photoelectron spectroscopy and DFT calculations

Xue Wu,[1] Xiaoqing Liang,[1] Qiuying Du,[1] Jijun Zhao,[1]* Maodu Chen,[1] Miao Lin,[2] Jiashuai Wang,[2] Guangjia Yin,[2] Lei Ma,[2]* R. Bruce King,[3] Bernd von. Issendorff [4]

[1] Key Laboratory of Materials Modification by Laser, Ion and Electron Beams (Dalian University of Technology), Ministry of Education, Dalian 116024, China  
[2] Tianjin International Center of Nanoparticles and Nanosystems, Tianjin University, Tianjin, 300072, China  
[3] Department of Chemistry and Center for Computational Chemistry, University of Georgia, Athens, Georgia 30602, USA  
[4] Department of Physics and FMF, University of Freiburg, D- 79104 Freiburg, Germany

E-mail: zhaojj@dlut.edu.cn and lei.ma@tju.edu.cn

**Abstract**

Size-selected anionic silicon clusters, $Si_n^-$ (*n*=14–20), have been investigated by photoelectron spectroscopy and density functional theory (DFT) calculations. Low-energy structures of the clusters are globally searched for by using a genetic algorithm based on DFT calculations. The electronic density of states and VDEs have been simulated by using ten DFT functionals and compared to the experimental results. We systematically evaluated the DFT functionals for the calculation of the energetics of silicon clusters. CCSD(T) single-point energies based on MP2 optimized geometries for selected isomers of $Si_n^-$ are also used as benchmark for the energy sequence. The HSE06 functional with aug-cc-pVDZ basis set is found to show the best performance. Our global minimum search corroborates that most of the lowest-energy structures of $Si_n^-$ (*n*=14–20) clusters can be derived from assembling tricapped trigonal prisms (TTP) in various ways. For most sizes previous structures are confirmed, whereas for $Si_{20}^-$ a new structure has been found.

## 1. Introduction

Silicon is the most important material in modern microelectronics and semiconductor industries. In this respect, the continuously increasing miniaturization of silicon-based transistors requires a deeper understanding of fundamental physical and chemical properties of silicon clusters and nanostructures.[1] Since silicon clusters favor $sp^3$-like covalent bonding, surface reconstruction reducing the number of dangling bonds is usually strong and can lead to distinctive cluster structures with electronic properties very different from those of the bulk diamond phase. Medium-sized silicon clusters are particularly interesting since they represent the key intermediates in the transition of silicon from molecular to bulk states.

During the past three decades, an enormous amount of experimental[2-11] and theoretical[12-22] efforts have been devoted to the understanding of the physics behind the unique properties of silicon clusters. Much attention has been focused on unveiling the general growth behavior of their structures especially for the small to medium-sized silicon clusters up to 30 atoms.[4-9, 12-26] Spectroscopic experiments using photoelectron, Raman, and infrared techniques have uncovered the geometric sequence of TTP based morphologies for silicon clusters of 9 to 26 atoms.[5-9, 13] Ion mobility measurements have also provided key evidence for the prolate-to-spherical shape transition of medium-sized silicon clusters. However, the TTP units are not ubiquitously found in the prolate species,[13, 23] thereby indicating the existence of new geometric building blocks.

For *ab initio* calculations of clusters, high levels of theory are always desirable. For instance, Tam et al. performed CCSD(T) computations on the $Si_3$ cluster and its derivatives containing an attached cation ($H^+$, $Li^+$, $Na^+$, $K^+$).[27] However, considering the extremely high computational requirements for CCSD(T) methods, such an approach usually is not practical for even medium-sized clusters. DFT methods with a relevant approximation of exchange-correlation interaction provide a compromise between computational cost and accuracy. So far, numerous DFT based theoretical calculations have been performed to explore the structural evolution of silicon clusters. Ho et al.[14] discovered that many low-energy isomers for $Si_{12-18}$ clusters contain the TTP motif of a $Si_9$ subunit by using a genetic algorithm global energy minimum search program with local density approximation (LDA) calculations. Moreover, with the PBE functional, Goedecker et al. found the TTP motif in $Si_{16-19}$ clusters using the dual minima hopping method to determine the global minima of the potential energy surfaces.[15] Rata et al.[16] found a six/six motif (i.e., a sixfold-puckered hexagonal $Si_6$ ring plus a six-atom tetragonal bipyramid $Si_6$) and a transition from TTP to a six/six motif at n = 19, under the framework of GGA using PBE and PWB88. However, Yoo's calculations show that the six/six motif is favored energetically over TTP at the B3LYP calculations already in the size range from 16 to 20.[17-18]

Although silicon clusters have been studied for years, the predicted lowest-energy geometries of silicon clusters remain controversial since the calculated energetic sequence of isomer structures depend highly upon the theoretical approach used. For instance, Liu et al. found that the "prolate" and "spherical" isomers become practically isoenergetic within GGA (PWB and BLYP functionals), where LDA clearly prefers energetically the latter.[12] Yoo et al. concluded that the motif transition from TTP-to-six/six heavily relies on the exchange-correlation functional used. Namely the B3LYP method slightly favors the six/six motif, whereas the PBE0 method favors the TTP motif. [17] For the $Si_{21-38}$ systems, nearly spherical endohedral fullerene structures are energetically favorable using the PBE functional, while the "Y-shaped three-arm" motif dominates when the BLYP functional is used.[28-29] More specifically, the lowest-energy silicon cluster isomers predicted by *ab initio* molecular orbital calculations at HF, MP2, MP3, MP4, CCSD, and CCSD(T) levels show high sensitivity toward the treatment of electron correlation.[19] Mitas et al. also concluded that electron correlation has an important impact on the overall stability of silicon cluster isomers.[26] Thus high-accuracy methods are necessary to uncover the true energetic ordering of isomers.

By revealing the fingerprint of their electronic structure photoelectron spectroscopy (PES) plays a crucial role in the identification of the ground-state configurations of anionic atomic clusters.[2-3, 8, 30-35] At a given cluster size, matching the simulated and measured PES results is a more sensitive probe to distinguish structurally distinct isomers than just the comparison of ionization potentials or electron affinities.[9] We report here studies on anionic $Si_n^-$ (n=14–20) clusters for which a large database of low-energy structures is available. We use the results of PES performed at low temperatures to evaluate systematically the performance of ten different DFT functionals in predicting the geometric and electronic state of silicon clusters. In addition, CCSD(T) results for a few isomers of $Si_{14}^-$ cluster are used as benchmark for the energy sequence of the isomers. Moreover, we evaluate the effect of basis set size on the simulation results. We conclude that the HSE06 or PBE0 functional combined with the aug-cc-pVDZ basis set provides the most accurate description of the structural and electronic properties of silicon clusters anions.

## 2. Experimental and Theoretical Methods

### 2.1 Experimental Methods

Silicon clusters were produced in a magnetron gas aggregation cluster source. Silicon was sputtered from a 2-inch target into a mixture of helium and argon buffer gas (roughly 3:1) with a total pressure of around 0.5 mbar inside a liquid nitrogen cooled aggregation tube. Under these conditions silicon clusters of various sizes are formed. The high density of charge carriers produced by the magnetron discharge leads to an effective charging of the clusters forming both anions and cations. The clusters were expanded with the buffer gas through an adjustable iris into vacuum. A radio-frequency (RF) octupole guided the cluster anions into the next chamber where they were fed into an RF 12-pole cryogenic ion trap. In this trap, which is cooled to 80 K, the clusters were thermalized by collisions with precooled helium buffer gas with a pressure of about $10^{-3}$ mbar. Bunches of cluster anions were then extracted from the trap and entrained into a high resolution, double reflection time-of-flight mass spectrometer, where a multiwire mass gate located at the focal point of the first reflector selected a specific mass with a resolution of about m/Δm=2000. The size-selected clusters were then reflected and rebunched again by the second reflector, and decelerated by a pulsed electric field. They then entered into the interaction region of a magnetic bottle time-of-flight photoelectron spectrometer, where they were irradiated by a laser pulse from a KrF excimer laser to acquire the photoelectron spectra. Typically, photoelectron spectra were averaged over 30000 laser shots at a repetition rate of 100 Hz. The spectrometer was calibrated by measuring the known photoelectron spectrum of platinum, thereby leading to an error of the measured binding energies of less than 30 meV.

*2.2 Theoretical Methods*

To find the most suitable exchange-correlation functional for describing silicon clusters within DFT, here we systematically evaluated a variety of common functionals including B3LYP,[36] cam-B3LYP,[37] B3PW91,[36] PBE,[38] PBE0,[39] HSE06,[40] TPSS,[41] TPSSh,[41] M06,[42] and M06-2X.[42] An aug-cc-pVDZ basis set was chosen for the DFT calculations.[43] Zero-point-energy (ZPE) corrections were included in the final energy of each cluster isomer for all calculations. CCSD(T) (combined with aug-cc-pVDZ basis set) single-point energies based on MP2 optimized geometries for selected isomers of $Si_{14}^-$ are also used as benchmark for the energy sequence. To compare with experimental PES data, the vertical detachment energies (VDEs) of the isomers were calculated using the above-mentioned functionals combined with aug-cc-pVDZ basis set. By definition, the VDE is the energy required to remove an electron from the highest occupied molecular orbital (HOMO) without relaxing the atomic configuration, and it corresponds to the first peak maxima of PES. On the other hand, the adiabatic detachment energies (ADEs) is the difference in total energy between the anionic and neutral clusters in their optimized geometries, which corresponds to the leading edge of the first peak of PES. The calculated DOS was then globally shifted in order to align the binding energy of the HOMO with the theoretical VDE value.[44-45]

In order to evaluate further the effect of the basis set size, the VDE of the lowest-energy structures of $Si_n^-$ ($n$=14–20) were computed using several different basis sets, such as 6-31G(d), 6-311G(d), 6-311+G(d), aug-cc-pVDZ, and aug-cc-pVTZ, combined with the HSE06 functional. Although many possible spin multiplicities were considered, doublets for $Si_n^-$ anion and singlets for neutral $Si_n$ are the lowest-lying spin state. All of these DFT calculations were performed using the Gaussian09 package.[46]

The structures of the $Si_n^-$ ($n$=14–20) cluster isomers were searched independently through a homemade unbiased comprehensive genetic algorithm (CGA) code[47] which is incorporated into the DMol$^3$ program for energy calculations.[48] The validity and efficiency of this CGA-DFT scheme have been well documented in a series of recent studies on Si,[21] B,[49] Pt-Sn,[50] V-Si,[51-52] B-Si,[53] and Fe-Ge[54] clusters. For each cluster size, a few independent GA searches were performed with different presumed symmetries, including the $C_1$, $C_2$, $C_3$, and $C_s$ point groups. With each specific symmetry constraint (no constraint for $C_1$), every GA search ran at least 3000 iterations and retained 16 members in the population. The mutation ratio was set as 40% to ensure the diversity of the population. The description of CGA in detail can be found in a recent review paper.[47]

**3. Results and Discussion**

*3.1 Validation of Density Functional and Basis Set*

The potential energy surfaces of anionic $Si_n^-$ ($n$=14–20) clusters were explored in an unbiased manner using the CGA-DFT scheme. For each size, the putative ground state configuration along with several important low-energy isomers was obtained and optimized further using ten different density functionals combined with the aug-cc-pVDZ basis set. The low-energy isomers of the $Si_n^-$ ($n$=14–20) clusters are represented as $Si_n^-$-I, $Si_n^-$-II, $Si_n^-$-III, $Si_n^-$-IV, $Si_n^-$-V, $Si_n^-$-VI, and $Si_n^-$-VII (Figure S1 of the Supporting Information). Those isomers include previously reported structures.[9, 22] Their relative energy differences are summarized in Table S1. The photoelectron spectra of every isomer are then simulated and depicted in Figure S1. The experimentally measured spectra of the $Si_n^-$ ($n$=14–20) clusters were also used as references to evaluate the validity of the DFT functionals. In order to facilitate the comparison, we integrated the simulated spectra and the measured ones in Figure S2 (for a photon energy of 5.0 eV) and Figure S3 (for a photon energy of 6.40 eV) of the Supporting Information. The lowest-energy isomers that match best with the experimental PES are shown in Figure 1.

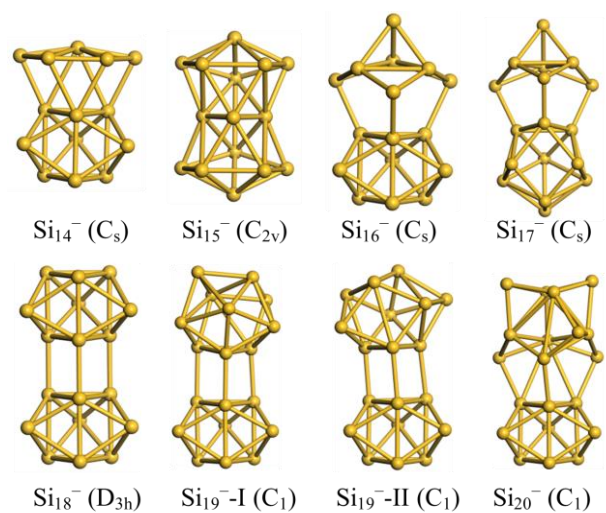

**Figure 1**. Low-energy structures of $Si_n^-$ ($n$=14–20) clusters found using HSE06/aug-cc-pVDZ. The cluster symmetry is given in parentheses.

**Table 1**. Experimental and theoretical vertical detachment energies (VDEs) of $Si_n^-$ ($n$=14–20) clusters. The theoretical VDEs were calculated by employing B3LYP, cam-B3LYP, B3PW91, PBE, PBE0, HSE06, TPSS, TPSSh, M06, and M06-2X functional, respectively. The *SD* is the standard deviation of VDE for a given functional. *SD* is the square root of its variance which is the average of the squared differences from the mean. All energies are given in eV. The uncertainties in the last digits of the experimental VDEs are shown in parentheses.

| n  | B3LYP | cam-B3LYP | B3PW91 | PBE   | PBE0  | HSE06 | TPSS  | TPSSh | M06   | M06-2X | Expt.   |
|----|-------|-----------|--------|-------|-------|-------|-------|-------|-------|--------|---------|
| 14 | 2.67  | 2.69      | 2.69   | 3.37  | 3.56  | 3.54  | 3.40  | 3.48  | 3.36  | 3.44   | 3.50(6) |
| 15 | 3.60  | 3.14      | 3.63   | 3.39  | 3.64  | 3.61  | 3.41  | 3.51  | 3.46  | 3.44   | 3.56(6) |
| 16 | 3.82  | 4.00      | 3.87   | 3.66  | 3.87  | 3.85  | 3.67  | 3.76  | 3.64  | 3.35   | 3.92(6) |
| 17 | 3.52  | 3.65      | 3.59   | 3.42  | 3.60  | 3.58  | 3.44  | 3.51  | 3.53  | 2.91   | 3.55(6) |
| 18 | 3.42  | 3.64      | 3.51   | 3.32  | 3.53  | 3.50  | 3.34  | 3.42  | 3.27  | 3.38   | 3.58(6) |
| 19 | 3.64  | 3.68      | 3.38   | 3.21  | 3.43  | 3.37  | 3.22  | 3.30  | 3.68  | 3.49   | 3.36(6) |
| 20 | 3.44  | 3.58      | 3.54   | 3.29  | 3.56  | 3.54  | 3.30  | 3.95  | 3.86  | 4.07   | 3.58(6) |
| SD | 0.341 | 0.368     | 0.310  | 0.210 | 0.055 | 0.052 | 0.193 | 0.168 | 0.235 | 0.388  |         |

In order to conduct a quantitative evaluation, we calculated the VDEs of the ground-state configurations optimized using the ten different functionals. It is worth pointing out that the geometries were relaxed separately for each functional, and the bond lengths changed within ~0.03 Å. The results are shown in Table 1 together with the experimental values. The minimum standard deviation (*SD*) of the differences between the calculated and measured VDEs of the ground state at each cluster size is used as the criterion to determine the functional with the best match. For the $Si_n^-$ ($n$=14–20) clusters, the standard deviation of the VDEs for different functionals are obtained as: B3LYP (0.341 eV), cam-B3LYP (0.368 eV), B3PW91 (0.310 eV), PBE (0.210 eV), PBE0 (0.055 eV), HSE06 (0.052 eV), TPSS (0.193 eV), TPSSh (0.168 eV), M06 (0.235 eV), and M06-2X (0.388 eV). These data indicate that the PBE0 and HSE06 methods with the lowest *SD* values (down to about 0.05 eV) are the most accurate in describing the electronic structure of silicon clusters. Therefore, the VDEs of the lowest-energy structures calculated by the PBE0 and HSE06 methods are selected to be compared with the experimental data as shown in Figure 2.

PBE0 is a parameter-free functional obtained by combination of the PBE functional with a predefined amount of exact exchange.[39, 55] The non-empirical derivation of the PBE0 functional makes it widely applicable for description of isolated molecules as well as condensed matter.[39] The use of a screened Coulomb potential for short-range Hartree–Fock (HF) exchange enables HSE06 to be a functional for fast and accurate hybrid calculations.[56] Application of the high accuracy of the screened Coulomb potential hybrid reveals all physically relevant properties of the full HF exchange, along with its computational advantages. This makes HSE06 particularly powerful for calculation of both large molecules and periodic systems.

Note that the VDEs from the HSE06 calculations are even slightly better than those obtained by using the PBE0 hybrid functional.

Using CCSD(T) results as benchmark, we considered five isomers of $Si_{14}^-$ and compared their relative energies obtained by different methods, as summarized in Table 2. Among various functionals, only the HSE06, PBE0, and TPSSh methods reproduce the correct energy order of isomers obtained by CCSD(T) calculations, while the PBE0 and HSE06 methods perform slightly better.

In addition to PBE0 and HSE06, a meta-hybrid-GGA functional, TPSSh, also displays reasonably reliable performance. On the other hand, the B3LYP, cam-B3LYP, M06, and M06-2X functionals yield neither the correct energy sequence of the cluster isomers nor the right VDE values. Among these functionals, B3LYP underestimates the VDEs, but the corrected cam-B3LYP severely overestimates the VDEs; M06 largely underestimates the VDEs, and M06-2X give slightly better VDEs relative to M06. In summary, the PBE, TPSS, and B3PW91 functionals perform better than the B3LYP, cam-B3LYP, M06, and M06-2X functionals, but they still cannot give the correct energy order within the range of experimental error. Similar to PBE, the meta-GGA TPSS functionals underestimates the VDEs. B3PW91 behaves even worse regarding VDE predictions. Inspired by the previous work,[7] we also took LDA functionals into account, and found that the LDA gives similar energy sequence of the cluster isomers to PBE and TPSS functionals, but significantly overestimates VDE (3.91 eV).

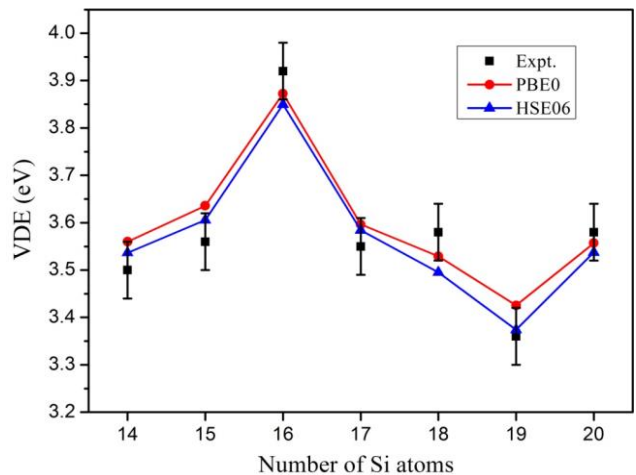

**Figure 2**. Vertical detachment energies (VDE) of the lowest-energy $Si_n^-$ structures as a function of cluster size. Black squares show experimental values with error bars; red circles show calculated results using the PBE0 functional; blue upper triangles show calculated results using the HSE06 functional. All computations were performed using the aug-cc-pVDZ basis set.

**Table 2**. Relative energies (eV) of five Si$_{14}^-$ isomers: Si$_{14}^-$-I, Si$_{14}^-$-II, Si$_{14}^-$-III, Si$_{14}^-$-IV, Si$_{14}^-$-V. All structures are fully relaxed with the respective methods, except that the CCSD(T) results are from a single-point energy calculation based on MP2 geometry optimization. The *SD* is the standard deviation of energy difference for a given functional. *SD* is the square root of its variance which is the average of the squared differences from the mean. All computations were conducted using the aug-cc-pVDZ basis set.

|  | Si$_{14}^-$-I | Si$_{14}^-$-II | Si$_{14}^-$-III | Si$_{14}^-$-IV | Si$_{14}^-$-V | *SD* |
|---|---|---|---|---|---|---|
| CCSD(T) | 0 | 0.11 | 0.17 | 0.44 | 0.59 |  |
| HSE06 | 0 | 0.07 | 0.17 | 0.32 | 0.49 | 0.072 |
| PBE0 | 0 | 0.07 | 0.17 | 0.33 | 0.50 | 0.066 |
| TPSSh | 0 | 0.12 | 0.13 | 0.32 | 0.49 | 0.072 |
| PBE | 0 | 0.15 | 0.11 | 0.34 | 0.53 | 0.061 |
| TPSS | 0 | 0.15 | 0.11 | 0.32 | 0.50 | 0.074 |
| B3LYP | 0.32 | 0 | 0.54 | 0.52 | 0.52 | 0.229 |
| cam-B3LYP | 0.42 | 0 | 0.70 | 0.58 | 0.57 | 0.313 |
| B3PW91 | 0.02 | 0 | 0.20 | 0.32 | 0.45 | 0.097 |
| M06 | 0 | 0.73 | 0.25 | 0.44 | 0.90 | 0.312 |
| M06-2X | 0 | 0.45 | 0.10 | 0.39 | 0.57 | 0.157 |
| LDA | 0 | 0.24 | 0.07 | 0.41 | 0.72 | 0.095 |

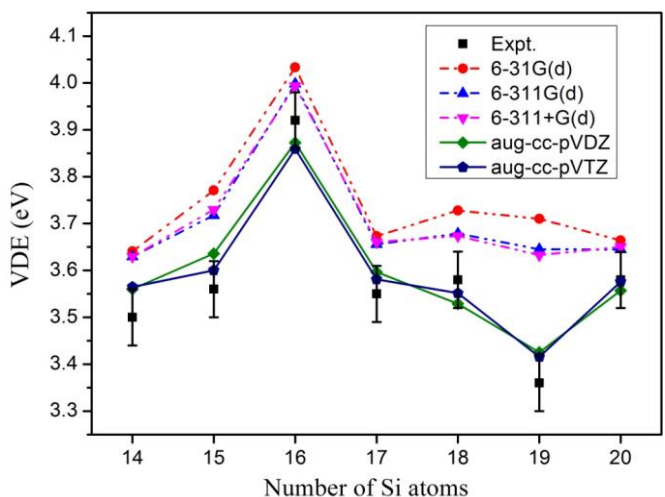

**Figure 3**. Vertical detachment energies (VDE) of the lowest-energy Si$_n^-$ structures as a function of cluster size. Black squares: experimental results with error bars; red circles: 6-31G(d); blue upper triangles: 6-311G(d); magenta lower triangles: 6-311+G(d); olive diamonds: aug-cc-pVDZ; navy pentagons: aug-cc-pVTZ. All computations were conducted by using the HSE06 functional.

The excellent performance of the PBE0 and HSE06 functionals in predicting the energy sequence of isomers and VDEs of ground state silicon clusters suggests that these two functionals should be a reliable choice for DFT calculations of large silicon clusters and nanostructures. In order to evaluate further the reliability of these two functionals, the electronic band structure of bulk silicon solid in the diamond phase was calculated using the CASTEP program,[57] which is based on the plane-wave pseudopotential technique with a cutoff energy of 600 eV. The calculated indirect band gap of bulk silicon is 1.15 eV using the HSE06 method, which agrees excellently with the experimental value of 1.17 eV at 0 K.[58] In contrast, PBE0 significantly overestimates the band gap by 0.74 eV. Bearing the performance of bulk silicon in mind, we conclude that HSE06 is the best functional out of the ten functionals mentioned above for distinguishing the isomer structures of silicon clusters and nanostructures as well as for description of their electronic properties.

We also discuss the effect of basis set by calculating the VDEs of the lowest-energy structures of Si$_n^-$ ($n$=14–20) clusters with the HSE06 functional. The different sized basis sets 6-31G(d), 6-311G(d), 6-311+G(d), aug-cc-pVDZ, and aug-cc-pVTZ were used. The results are shown in Figure 3. Clearly, 6-31G(d), 6-311G(d), and 6-311+G(d) are insufficient to describe silicon clusters accurately, since they overestimate considerably the VDEs in general compared with the other basis sets considered. Both the aug-cc-pVDZ and aug-cc-pVTZ basis sets perform well. However, aug-cc-pVTZ is computationally much more demanding, requiring almost ten times more computational power than aug-cc-pVDZ. Therefore, we conclude that the aug-cc-pVDZ basis sets combined with HSE06 or PBE0 are the most reasonable and effective methods to describe silicon clusters.

*3.2 Lowest-energy Structures*

Here, we discuss the low-energy configurations of Si$_n^-$ ($n$=14–20) clusters using the HSE06 method (Figure 1). The $D_{3h}$ TTP and the $D_{4d}$ bicapped square antiprism (BSA) are found to be important building blocks for the silicon clusters containing at least 14 vertices (Figure 1). In fact these deltahedra are also found in the stable borane dianions B$_n$H$_n^{2-}$ ($n$=9, 10)[59]. They have only degree 4 and 5 vertices, where the degree of a vertex is the number of edges meeting at the vertex in question (Figure S4). Other significant building blocks are the deformed TTP and deformed BSA with a single degree 6 vertex. The deformed ones are related to the nondeformed ones by a single diamond-square-diamond rearrangement.

In general, like earlier studies we found that the lowest-energy $Si_n^-$ ($n$=14–20) clusters are always formed by joining two smaller structural units. There are two possible types of junctions, namely face-sharing and single linkage through external Si–Si bond formation. The TTP building block is particularly prevalent in most of the low-energy structures. A 7-vertex $Si_7$ unit, best described as a triply edge-bridged tetrahedron, is also found in a few of the lowest-energy structures.

The low-energy structures of the silicon clusters $Si_n^-$ ($n$=14–20) can be summarized as follows: $Si_{14}^-$-I is a face-sharing fusion of a TTP with a bicapped octahedron, which agrees well with the DFT study by Shvartsburg et al..[13] The metastable $C_{2v}$ $Si_{14}^-$-II isomer, which is 0.07 eV higher in energy, can be viewed as a distorted tube. The $Si_{14}^-$-III and $Si_{14}^-$-IV isomers, which lie 0.17 and 0.32 eV, respectively, in energy above $Si_{14}^-$-I, are both constructed from a TTP unit. The $C_{2v}$ $Si_{15}^-$-I structure consists of a face-sharing fusion of two deformed TTPs. This differs from the prediction of Shvartsburg et al.,[13] but agrees with the results of Rata et al..[16] Note that the mobility calculated by Rata et al. is much closer to the experimental measurement than that calculated by Shvartsburg et al.. In our calculation, $Si_{15}^-$-III (see Figure S1) has the same structure as the ground state predicted by Shvartsburg et al., but in fact is energetically less stable than $Si_{15}^-$-I by 0.23 eV.

The $C_s$ lowest-energy $Si_{16}^-$ structure consists of a TTP linked by three external Si–Si bonds to a triply edge-bridged tetrahedron. Isomers $Si_{16}^-$-II and $Si_{16}^-$-III are energetically less stable than $Si_{16}^-$-I by 0.17 and 0.18 eV, respectively. The $C_{3v}$ metastable $Si_{16}^-$-II structure has a TTP linked to a tetrahedron by Si–Si bonds with its faces capped by the three remaining silicon atoms.

The top half of the $Si_{17}^-$-I structure keeping with $Si_{16}^-$-I has $C_s$ symmetry. $Si_{17}^-$-I consists of a BSA linked by three external Si–Si bonds to a triply edge-bridged tetrahedron. The other isomers lie at least 0.20 eV in energy above $Si_{17}^-$-I. Among them, the $Si_{17}^-$-II structure can be derived from the $Si_{16}^-$-II structure by adding an edge-bridging Si atom to the tetrahedron. The $D_{3h}$ ground-state of $Si_{18}^-$ is formed by two TTPs linked by three external Si–Si bonds. This is also consistent with the result of Shvartsburg et al..[13] The metastable $Si_{18}^-$-II isomer, which is 0.21 eV higher in energy, has a TTP linked by three external Si–Si bonds to a deformed TTP.

Two nearly degenerate structures were found for $Si_{19}^-$ with only a small energy difference of 0.01 eV. The lowest energy $Si_{19}^-$-I structure has a deformed BSA linked to a TTP by three Si–Si bonds. The $Si_{19}^-$-II isomer has slightly higher energy but is structurally very similar to $Si_{19}^-$-I except that the BSA is less deformed. The configuration of $Si_{19}^-$-I agrees well with the one found by Rata et al..[16]

For $Si_{20}^-$, a new ground state has been found having a TTP linked by five Si–Si bonds to a not-readily-recognizable $Si_{11}$ unit. This $Si_{11}$ unit can also be properly described as an 8-vertex polyhedron having one pentagonal and one tetragonal face with three edges bridged by silicon atoms. Compared to experimental results, the VDE error of $Si_{20}^-$ has been dramatically improved over the previous structures.[20] In addition, the simulated photoelectron spectrum of the new ground state configuration agrees well with the experimentally measured spectrum as discussed later in the next subsection. The ground state structure of $Si_{20}$ reported by Bai et al.,[22] is indeed the $Si_{20}^-$-II (see Figure S1). This structure containing two fused hexagonal rings, is completely different from that of $Si_{20}^-$-I and its energy is slightly higher (0.08 eV, HSE06/aug-cc-pVDZ) than $Si_{20}^-$-I. Our new $Si_{20}^-$ anion is found to be lower in energy than all previously reported geometries (Table S1).[9, 22]

*3.3 Photoelectron Spectra*

In Figure 4, the photoelectron spectra of the $Si_n^-$ ($n$=14–20) clusters recorded with 248 nm photons are compared with the simulated photoelectron spectra of the lowest-energy isomers (HSE06/aug-cc-pVDZ). Figure S5 also compares the simulated spectra of four other low-energy isomers with the measured ones. Increasing the cluster size leads to a rapid increase of the density of electronic states near the Fermi level thereby broadening the features in the measured PES. This makes definitive identification more difficult. We therefore chose as a benchmark study the intermediate size range ($n$=14–20) of silicon clusters. In this work, the comparison between the PES peaks, VDE and ADE values reported in previous studies are also presented (Table 3).

From the measured photoelectron spectra, many spectral features can be discerned. The $Si_{14}^-$ spectrum reveals a ground-state (X) with a VDE of 3.50 eV as well as a broad band (A) above 4.3 eV, with a shoulder starting at 4.05 eV. The simulated photoelectron spectrum matches very well with the measurement with only a negligible shift within the range of experimental error.

The X and A features of $Si_{15}^-$ overlap, centered at 3.56 and 3.95 eV, respectively. The broad B band starts at a binding energy of 4.4 eV. From Figure S3, the two peaks of the double peak at ~3.5 eV are too close for most functionals, and the minimum at ~5eV is practically absent for all of them. We were able to estimate binding energies for the X and A bands of $Si_{16}^-$ of 3.92 and 4.15 eV, respectively. There seems to be a small contribution from an additional isomer with a VDE of 3.68 eV.

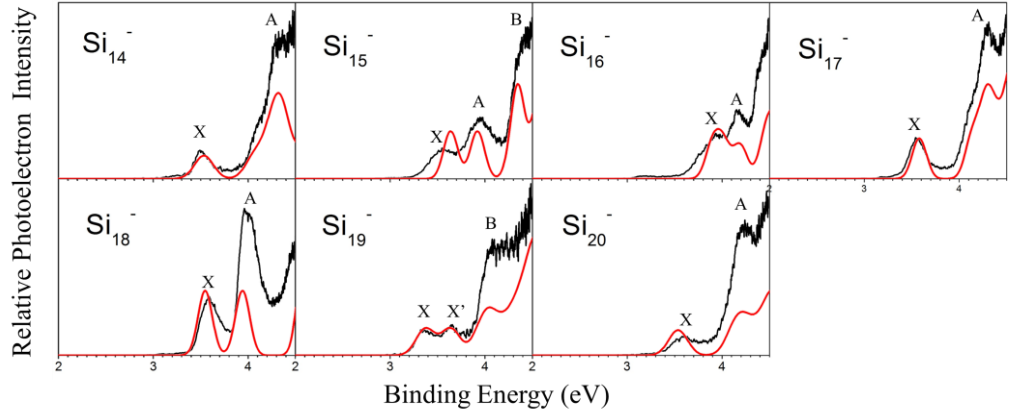

**Figure 4**. Photoelectron spectra of cold (T = 80 K) silicon cluster anions, measured at a photon energy of 4.99 eV (thick dark lines) and combined with the simulated photoelectron spectra from HSE06/aug-cc-pVDZ calculations for the lowest-energy structure (thin light red lines). For $n$=19, the simulated photoelectron spectra is a sum of $Si_{19}^-$-I and $Si_{19}^-$-II. The simulations were conducted by fitting the distribution of the Kohn-Sham energies with unit-area Gaussian functions using 0.06 eV broadening.

**Table 3**. Comparison of the theoretical (theo.) VDE's and ADE's with the experimentally measured (expt.) VDE's and ADE's in eV for the $Si_n^-$ ($n$=14–20) clusters.

| n | Feature | VDE(theo.) | VDE (expt.) | | | ADE(theo.) | ADE(expt.) | |
|---|---------|------------|-------------|---|---|------------|------------|---|
| | | | This work | a ref. | b ref. | | This work | a ref. |
| 14 | X | 3.54 | 3.50 | 3.45 | | 3.45 | 3.40 | 3.30 |
| | A | 4.32 | 4.30 | 4.28 | | | | |
| 15 | X | 3.61 | 3.56 | 3.55 | | 3.12 | 3.20 | 3.10 |
| | A | 3.82 | 3.83 | 3.94 | | | | |
| | B | 4.32 | 4.40 | 4.59 | | | | |
| 16 | X | 3.85 | 3.92 | 3.99 | | 3.47 | 3.52 | 3.45 |
| | A | 4.18 | 4.15 | 4.52 | | | | |
| 17 | X | 3.58 | 3.55 | 3.50 | | 3.33 | 3.35 | 3.20 |
| | A | 4.30 | 4.28 | 4.35 | | | | |
| 18 | X | 3.50 | 3.58 | 3.52 | | 3.33 | 3.38 | 3.25 |
| | A | 3.90 | 3.99 | 3.93 | | | | |
| 19 | X | 3.37 | 3.36 | 3.40 | | 3.30 | | |
| | X' | 3.62 | 3.62 | | | 3.16 | 3.18 | 3.10 |
| | A | 4.06 | 4.08 | 4.10 | | | | |
| 20 | X | 3.54 | 3.58 | 3.45 | 3.57 | 3.36 | 3.35 | 3.20 |
| | A | 4.21 | 4.22 | 4.12 | 4.20 | | | |

<sup>a</sup> ref. [7]  <sup>b</sup> ref. [22]

The $Si_{17}^-$ spectrum has a well separated X band centered at 3.55 eV, and an A band centered at 4.28 eV, with a shoulder indicating another state at about 4.1 eV. The simulated photoelectron spectra agree well with the measured PES of $Si_{17}^-$. The X band of $Si_{18}^-$ is centered at 3.58 eV with a reasonably sharp onset. Band A has a binding energy of 3.99 eV. The simulated photoelectron spectrum is systematically shifted by 0.03 eV to lower energies as compared with the experimental spectrum.

The spectrum of $Si_{19}^-$ has two overlapping featured peaks (X and X') centered at 3.36 and 3.65 eV. The simulated photoelectron spectrum of the lowest-energy structure of $Si_{19}^-$-I cannot match well since the second feature peak is missing. Because of their small energy difference, the $Si_{19}^-$-I and $Si_{19}^-$-II isomers can be assumed to be both present in the experiment. Therefore we can reasonably interpret the measured spectra as arising from a combination of the $Si_{19}^-$-I and $Si_{19}^-$-II spectra, that is, $Si_{19}^-$-I contributes solely to the lowest energy band (X) whereas $Si_{19}^-$-II mainly contributes to the X' band. The spectrum of $Si_{20}^-$ has a resolved band X centered at 3.58 eV, followed by a prominent peak centered at 4.22 eV. The simulated photoelectron spectrum of the $Si_{20}^-$

-I agrees rather well with the experimental spectra. While the metastable state structure ($Si_{20}^-$-II) reveals a single peak centered at 4.10 eV which is disagrees with the experimental spectra, as is shown in Figure S5. Hence, we suggest that isomer $Si_{20}^-$-I is the most probable structure detected in the experiment.

## 4. Conclusions

A systematic investigation on the low-energy structures of the $Si_n^-$ ($n$=14–20) clusters has been performed using DFT calculations combined with high resolution photoelectron spectroscopy data. We carefully evaluated the performance of ten different density functionals and several basis sets of different sizes for the electronic structure calculations and ground-state searching of $Si_n^-$ ($n$=14–20) clusters using experimental photoelectron spectra as well as CCSD(T) calculations of $Si_{14}^-$ isomers as a basis of comparison. The results show that HSE06 and PBE0 are the most reliable functionals for DFT calculations of silicon clusters and related nanostructures. The B3LYP, cam-B3LYP, M06, and M06-2X functionals neither yield the correct energy sequence of the cluster isomers nor give the right VDEs. The PBE, TPSS, TPSSh, and B3PW91 functionals all perform reasonably well. However they still cannot yield the correct energies within the range of experimental error. The HSE06 functional is recommended for the accurate description of the electronic structure of silicon clusters.

As for the basis set, the sizes of the 6-31G(d), 6-311G(d), and 6-311+G(d) are insufficient to describe silicon clusters accurately, since they overestimate the VDEs. The larger aug-cc-pVDZ and aug-cc-pVTZ basis sets provide accurate results. Considering the computational efficiency, the HSE06 functionals with the aug-cc-pVDZ basis set using ZPE corrections are recommended as providing an optimum combination of accuracy and cost.

Applying this method to the silicon cluster anions, the ground state structures from our global search are found to reproduce the previously reported ones. In addition, we have also found a new $Si_{20}^-$ geometry having lower total energy than all previously known structures. The calculated photoelectron spectrum of the new $Si_{20}^-$ structure is in good agreement with experiment. Finally, the $Si_n^-$ ($n$=14–20) structures reveal remarkable structural motifs in the growth habit of prolate Si clusters besides the well-known TTP units. By comparing the performance of commonly used functionals for differentiating isomers of silicon clusters, we demonstrate here that great care should be taken regarding the choice of exchange-correlation functional for the investigation of silicon nanomaterials.

## Supporting Information

The structures and energetic data of the low-lying isomers of $Si_n^-$ ($n$=14–20) clusters. Photoelectron spectra of silicon cluster anions, measured at a photon energy of 4.99 and 6.40 eV, respectively, in comparison with the simulated photoelectron spectra for the lowest-energy structure, and photoelectron spectra (4.99 eV) compared with the simulated photoelectron spectra for five low-energy isomers of every size using the HSE06 functional and aug-cc-pVDZ basis set. Coordinate files of the lowest-lying structures of $Si_n^-$ ($n$=14–20) clusters.

## Acknowledgements

This work was supported by the National Natural Science Foundation of China (11574040, 11604039), the Fundamental Research Funds for the Central Universities of China (DUT16-LAB01, DUT17LAB19), the key program of Natural Science Foundation of Tianjin (17JCZDJC30100), the Seed Program of Tianjin University (2017XZC-0090) and the Supercomputing Center of Dalian University of Technology.